\documentstyle[12pt]{article}
\input{mssymb.sty}

\title{Quantum orbits of R-matrix type}
\author{J.Donin\\
Departement of Mathematics, Bar-Ilan University,\\
 52900 Ramat-Gan, Israel\\
D.Gurevich\\
Centre de Math\'ematiques, Ecole Polytechnique,\\
 F-91128 Palaiseau, France}
\date{}

\begin{document}

\newtheorem{proposition}{Proposition}
\newtheorem{conjecture}{Conjecture}
\newtheorem{definition}{Definition}
\newtheorem{remark}{Remark}
\newcommand{\gggg}{\mbox{${\frak g}$}}
\newcommand{\bC}{\mbox{$\Bbb C$}}
\newcommand{\bR}{\mbox{$\Bbb R$}}
\newcommand {\co}{{\cal O}_x}
\newcommand {\cR}{{\cal R}}
\newcommand{\ug}{U(\gggg)}
\newcommand{\gug}{Gr\,U(\gggg)}
\newcommand{\us}{U(sl(2))}
\newcommand{\vv}{V^{\otimes 2}}
\newcommand{\uqs}{U_q(sl(2))}
\newcommand{\oI}{\overline{I}}
\newcommand{\uq}{U_q(\gggg)}
\newcommand{\sss}{SL(n)/S(L(n-1)\times L(1))}
\newcommand{\St}{\tilde S}
\newcommand{\sn}{sl(n)}
\newcommand{\ppb}{P_{\beta}}
\newcommand{\pqb}{P^q_{\beta}}
\newcommand{\wkm}{\wedge_-^k(V)}
\newcommand{\wm}{\wedge_-(V)}
\newcommand{\vb}{V_{\beta}}
\newcommand{\va}{V_{\alpha}}
\newcommand{\vaa}{V_{2 \alpha}}
\newcommand{\vva}{V_{\alpha}^{\ot 2}}
\newcommand{\wpp}{\wedge_+(V)}
\newcommand{\wkp}{\wedge_+^k(V)}
\newcommand{\wkpm}{\wedge_{\pm}^k(V)}
\newcommand{\wpm}{\wedge_{\pm}(V)}
\newcommand{\ot}{\otime}
\newcommand {\mod}{\mbox{$U_q(\gggg)$-Mod}}
\newcommand {\mos}{\mbox{$U_q(sl(2))$-Mod}}
\newcommand{\oi}{\mbox{$\overline{I}$}}
\newcommand{\aq}{A^q}
\def\ot{\otimes}
\def\de{\Delta}
\newcommand{\ac}{A^q_c}
\def\sl{sl(2)}
 \def\qq{(q+q^{-1})}

\maketitle

\begin{abstract}
Given a simple Lie algebra $\gggg$, we consider the orbits in $\gggg^*$
which are
of R-matrix type, i.e., which possess a Poisson pencil
generated by the Kirillov-Kostant-Souriau bracket and the
so-called R-matrix bracket.
We call an  algebra quantizing the latter bracket a quantum orbit
of R-matrix type. We describe some orbits of this type explicitly and we
construct a quantization of the whole Poisson
 pencil on these orbits in a similar way.
The notions of q-deformed Lie brackets,
braided coadjoint vector fields
and tangent vector fields are discussed as well.
\end{abstract}

\section{Introduction}
It is well known that a quantum analog $Fun_q(G)$ of the function
space $Fun(G)$ on a simple Lie
group $G$ is the ``restricted dual'' of the quantum group $\uq$.
This algebra arises by a quantization of the Sklyanin-Drinfeld
 bracket\footnote{We prefer to use this name for the bracket
(introduced by E. Sklyanin and generalized by V. Drinfeld) usually
called the
 Poisson-Lie bracket
since the latter name is also used to designate the linear Poisson-Lie bracket
 on $\gggg^*$ arising from the Lie algebra structure of $\gggg$.},
which is defined on $G$ as follows
$$\{f,g\}_{SD}=\{f,g\}_l-\{f,g\}_r,\,\,\{f,g\}_{r,l}=\mu<\rho_{r,l}^{\ot 2}(R),
df\otimes dg>,\,\, f,g\in Fun(G),$$
where $\rho_r (\rho_l)$ is the representation of the corresponding
 Lie algebra $\gggg$ in
the space $Fun(G)$ by the left- (right-) invariant vector fields,
$\mu: Fun(G)^{\otimes 2}\to Fun(G)$ is the ordinary commutative multiplication
in $Fun(G)$ and $R$ is the R-matrix
\begin{equation}
R=\frac{1}{2}\sum_{\alpha \in \Delta_{+}}X_{\alpha}\wedge
X_{-\alpha} \in \wedge ^2 \gggg.
\end{equation}
Here $\{H_{\alpha},X_{\alpha},X_{-\alpha}\} $
is the Cartan--Weyl system in the Chevalley normalization and $\Delta_{+}$
is the set of positive roots of  $\gggg$. Here and below $Fun(M)$,
where $M$ is an algebraic variety in a vector space $V$, denotes the
restrictions of polynomials from $V$ to $M$.

Consider a homogeneous $G$-space $M=G/H$ where $H$ is a closed subgroup in $G$.
 A natural question arises: what is a
 quantum counterpart of the function space $Fun(M)$? The usual answer
is as follows. Assume that the algebra $Fun(M)$ is identified with the subspace
 $Fun(G)^H\subset Fun(G)$ of the functions invariant with respect to the
 right shift by any element of $H$. Assume that the bracket
$\{\,\,,\,\,\}_{SD}$ can be restricted to the space $Fun(G)^H$ (this means
that $\{f,g\}\in Fun(G)^H$ if $f,\,g \in Fun(G)^H$).
Then one considers this restricted bracket as a reduction of
 $\{\,\,,\,\,\}_{SD}$
on the space $M$, and the reduced bracket will be denoted
$\{\,\,,\,\,\}_{SD}^{red}$.

By a quantum analog of the space $Fun(M)$ one usually means the result
 of the quantization of the bracket $\{\,\,,\,\,\}_{SD}^{red}$
described in terms of ``quantum pairs''.
This essentially means
a pair of algebras $(A_1,\,\, A_2)\,\,,A_1=Fun_q(G),\,A_2=Fun_q(H)$
equipped with an epimorphism $P:A_1 \to A_2$ of Hopf algebras.
The space of elements
$a \in A_1$, ``invariant'' with respect to the coproduct
$$\Delta_P=(id\ot P)\Delta:A_1\to A_1\ot A_2,$$
where $\Delta$ is the comultiplication of $A_1$, i.e., such that
$\Delta_P (a)=a\otimes 1$, is then considered as a
quantum analog of the commutative algebra $Fun(M)$.

In the present paper we propose another way to treat the quantum counterpart of
 the space $Fun(M)$ for a certain class of $G$-homogeneous spaces $M=G/H$
which comprises the symmetric spaces
(in fact, we deal only with orbits in $\gggg^*$).
These spaces have a nice property: when the brackets $\{\,\,,\,\,\}_l$ and
$\{\,\,,\,\,\}_r$  are reduced (assuming
this reduction to be well-defined), they are both Poisson.

In fact we must only reduce the bracket $\{\,\,,\,\,\}_{r}$, while the
bracket $\{\,\,,\,\,\}_{l}$ is well defined on any
homogeneous $G$-space $M=G/H$.
Moreover, a bracket of such a type can be
assigned to any element $R\in \wedge ^2(\gggg)$ and any space $M$ equipped with
 a representation
$\rho:\gggg \to Vect(M)$ in a natural way:
$\{f,g\}_R=\mu<\rho^{\ot 2}(R),df\ot dg>$.
(In the sequel we use the notation $\{\,\,,\,\,\}_R$
for this bracket and call it the {\em R-matrix bracket}.)
However it is Poisson only under certain
conditions on $M$ and $R$.

Here we are dealing here with
the R-matrix bracket arising from the R-matrix (1) and we say that a manifold
$M$ equipped with such an R-matrix bracket is of {\em R-matrix type}
 if this bracket is Poisson.
Thus we can say that all symmetric $G$-spaces are of
R-matrix type. Nevertheless there exist other spaces of R-matrix type, e.g.,
the orbits of the highest weight vectors in any $G$-module $V$
(h.w. orbits for brevity).

Let us emphasize that on any orbit of
R-matrix type in $\gggg^*$ there exists
a Poisson pencil
\begin{equation}
\{\,\, ,\,\,\}_{a,b}=a\{\,\, ,\,\,\}_{KKS}+b\{\,\, ,\,\,\}_R
\end{equation}
generated by the Kirillov-Kostant-Souriau bracket
$\{\,\,,\,\,\}_{KKS}$ and the R-matrix bracket $\{\,\,,\,\,\}_R$,
since these brackets are always compatible.

Note that on any  orbit of R-matrix type in $\gggg^{*}$ there exist two
$G$-invariant Poisson brackets , namely, $\{\,\, ,\,\,\}_{KKS}$ and
the reduced Poisson bracket
$\{\,\, ,\,\,\}_{r}^{red}$ (assuming it to be well-defined).
The question of the  relation between these brackets arises. It is easy to see
that the bracket $\{\,\, ,\,\,\}_{r}$ is  identically equal to zero on the
h.w. orbit. In the case of
the symmetric orbits the brackets $\{\,\, ,\,\,\}_{KKS}$ and
$\{\,\, ,\,\,\}_{r}^{red}$ are proportional to each other.

This fact was first proved in \cite{DG2}
by a cohomological method.
Here, we give another, more direct, proof of this fact
using the aproach of \cite{KRR}
where the compatibility of the brackets
$\{\,\, ,\,\,\}_{KKS}$ and $\{\,\, ,\,\,\}_{SD}^{red}$
was proved. Note that the case of the symmetric orbits in $su(2)^*$
was also considered in \cite{SLW}.

Thus on a symmetric
orbit the bracket $\{\,\, ,\,\,\}_{SD}^{red}$ coincides  with a bracket
in  pencil (2). So quantizing
 the whole pencil (2) on a symmetric
orbit, we can get a quantization of
the bracket $\{\,\, ,\,\,\}_{SD}^{red}$ as a particular case.

Let us summarize the main difference between the above-mentioned aproach
to the quantization of homogeneous spaces based on quantum pairs
and our scheme.
We simultaneously quantize the Poisson pencil (2)
and get the quantum algebras for the brackets
$\{\,\,,\,\,\}_{SD}^{red}$ and $\{\,\, ,\,\,\}_{R}$ as particular cases.
(Remark that we consider the quantization of the latter bracket rather than
 of bracket $\{\,\, ,\,\,\}_{SD}^{red}$ as
a quantum analog of the space $Fun(\co)$.)

In the present paper we  realize this scheme for the
symmetric orbits of $\sss$ type in $sl(n)^*$
(the h.w. orbit in $\gggg^{*}$ belongs to closure of the family of the
above orbits and it is quantized as well).

The method of quantization consists in the following.
First, we describe the orbit under consideration
 by means of a system of algebraic equations and we
look for a q-analog of this system.
Second, we introduce a {\em q-Lie bracket}, which is a deformation of
the initial Lie bracket.
 Then the quantum algebra (denoted
$A_{h,q}^c$) is defined as a quotient algebra of the corresponding ``enveloping
algebra'' by  the ideal generated by the deformed system.

As a result we get a three parameter family $A_{h,q}^c$ of
associative algebras.
The parameter $q$  corresponds to the
quantization of the R-matrix bracket,
 the parameter $h$ is a factor introduced in the q-Lie bracket and
it correponds to the quantization of the KKS bracket, while $c$
labels the orbits (the case $c=0$ corresponds to the h. w.
orbit).

The space $\gggg$ equipped with the
 q-Lie bracket can be considered as a
``braided'' analog of a Lie algebra. We do not discuss here an axiomatic
definition of a ``braided Lie algebra'' (cf. \cite{G3}, \cite{DG3}
for a discussion). We only remark that
the deformed bracket is a very useful tool
to ``quantize'' certain objects of
ordinary analysis. In particular, by means of the q-Lie
bracket,  we introduce the  {\em braided
coadjoint vector fields}. We introduce also the {\em
braided tangent vector fields} as
elements of the  module generated over the quantum function algebra $A_{0,q}^c$
by the above fields.

The paper is organized as follows. In Section 2 we describe the
quantum algebras $A_{h,q}^c$. Section 3 is devoted to
the discussion of braided vector fields. An example of such fields
corresponding to the case $\gggg=sl(2)$ is considered  in
Section 5. In Section 4 we describe the quasiclassical limit of the
algebras $A_{h,q}^c$ and prove that on symmetric orbits the
brackets $\{\,\, ,\,\,\}_{KKS}$ and
$\{\,\, ,\,\,\}_{r}^{red}$ are proportional to each other.

\section{q-Lie brackets and quantum algebras}

Let us fix a simple Lie algebra $\gggg$ over the base field $k=\bR$
and the corresponding quantum
group $\uq$. Let $\gggg$-Mod ($\uq$-Mod) denote the tensor category of the
finite dimensional $\gggg-\, (\uq-)$
modules. Morphisms in these categories will be called $\gggg-\, (\uq-)$
morphisms.

For any linear space $V$ we denote $I_+\,\,(I_-)$
the subspace in $\vv$ of the symmetric (skew) tensors.

Let us consider the space $V=\gggg$ as an object  of the category
$\gggg$-Mod. Let $\vv=\oplus \vb $ be a decomposition of $\vv$
as a direct sum of isotypic $\gggg$-modules where
$\beta$ denotes the highest weight (h.w.) of the corresponding module.
Let $\alpha$ be h.w. of $\gggg$ as a $\gggg$-module.
Then in the above decomposition there is a trivial module $V_0$
($\beta=0$),
a  module $V_{2\alpha}$
of h.w. $\beta=2\alpha$, and an irreducible module $V_{\alpha}$ of
the h.w. $\alpha$ belonging to the space $I_-$. Note that for certain simple
algebras there exists in $\gggg^{\ot 2}$ another module of h.w. $\alpha$
which lies in $I_+$.

Note that the Lie bracket $[\,\,,\,\,]:\vv=\oplus V_{\beta}\to \gggg$
kills all modules except $V_{\alpha}$ and, when restricted to
$V_{\alpha}$, defines a $\gggg$-morphism.

We now introduce a deformed bracket in the following way. Let us
consider the space
$V$ as an object of the category $\uq$-Mod, i.e., define a left
representation
$\rho_q:\uq \to End(V)$ deforming the representation $\rho:\ug \to End(V)$ of
the enveloping algebra $\ug$ which  coincides on $\gggg$
with the representation
of $\gggg$ given by $\rho(X)=[X,\,.\,]$. By construction we assume that
$\rho_{q}\to\rho$ when $\rho\to 1$.

Let us define a representation $\rho^{\otimes 2}_q: \uq \to End(\vv)$ using
the coproduct $\Delta :\uq \to \uq^{\ot 2}$ and
decompose the space
$\vv$ as a direct sum $\oplus \vb^q$ in the category $\uq$-Mod. Hereafter,
 $\vb^q$ denotes the q-deformation of the $\gggg$-module $\vb$. We also
denote
by $I^q_+\,(I^q_-)$ the q-deformation of the subspace $I_+\,(I_-)$ defined
as follows.

Let us consider the universal R-matrix ${\cal R}$ corresponding to the
quantum group $\uq$. Let $S_{\rho}=\sigma \rho_{q}^{\ot 2}({\cal R}):\vv\to\vv$
be the ``braiding operator'' ($\sigma$ is the flip). Note that
$S_{\rho}\to \sigma$ when $q\to 1$. It is well known that
$S_{\rho}$ is diagonalizable. If $q-1$ is small enough each
eigenvalue of the operator  is close either to 1 or to $-1$.
Replacing the eigenvalues of the first type by 1 and those
of the second type by $-1$ we get another  operator (denoted $\tilde
S$) whose eigenspaces coincide with these of the operator $S_{\rho}$.
This operator is involutive $({\tilde S}^{2}=id)$, it is a morphism in
the category $\uq$-Mod but does not satisfy the QYBE.
Let us set $I^q_{\pm}=Im(id\pm \tilde S)$.

\begin{definition} We call q-Lie bracket the operator
$[\,\,,\,\,]_q:\vv\to V$ defined as
follows
$$[\,\,,\,\,]_q|_{\vb^q}=0\,\,for \,\, all\,\,
\vb^q\not=V_{\alpha}^q$$
and $ [\,\,,\,\,]_q:V_{\alpha}^q\to V$ is $\uq$-isomorphism.
\end{definition}

Note that $V_{\alpha}^q$ denotes the $\uq$-module of the h.w.
$\alpha$ which belongs to the space $I^{q}_{-}$.
Note also that the q-Lie bracket is defined up to a factor.

\begin{remark} In \cite{G1}, \cite{G2}, were introduced the notions
 of generalised (or $S$-) Lie algebra corresponding to an involutive
$(S^2=id)$ solution of the quantum Yang-Baxter equation, of
representation of such an algebra and the notion of corresponding
 enveloping algebra.
It is not difficult to define the category of representations of an S-Lie
algebra.

In the case under consideration the situation is inverse.
First we have a category. After that we introduce a q-Lie bracket
as a morphism in this category. Meanwhile we do not use any axioms of
Jacobi identity type.
We refer the reader to  \cite{DG3} where relations between
``braiding'' of Lie algebras and flatness of a
deformation of ``enveloping algebras''\footnote{Recall that
a family of associative algebras $A_h$
depending on a formal parameter $h$ is called a flat deformation of the initial
 algebra $A=A_0$ if there exists an isomorphism $A_h \to A[[h]]$ of
$k[[h]]$-modules.
Note that for any flat deformation of an associative algebra $A_0$
the skew-symmetrized linear term in $h$ of the
deformed product is a Poisson bracket.} are discussed (cf. also
\cite{M2} where another approach to ``braiding''  is realized without
connection with flatness of deformations).
\end{remark}

It seems very natural to introduce
 the braided enveloping algebra as the quotient algebra
\begin{equation}
T(V)/\{(Im(id-[\,\, ,\,\,]_q)I^q_-\}.
\end{equation}
Hereafter, $T(V)$ denotes the tensor algebra of the space $V$
and $\{I\}$ means the ideal generated  by a family $I\subset T(V)$.

However, this algebra is not a flat deformation
of its classical counterpart.
Moreover the deformation from $T(V)/\{I_-\}$ to
$T(V)/\{I^q_-\}$ is not flat
except in the case
$\gggg=sl(2)$. One can prove this fact by observing that the
R-matrix bracket is not Poisson on $\gggg^*$ if $\gggg\not=sl(2)$.

Nevertheless after this deformation is restricted to some
quotient algebra it becomes flat.
Consider an example. Let us introduce the following algebra
$T(V)/\{\oI^q_-\}$  where $\oI^q_-=\oplus_{\beta\not=2\alpha}V^q_{\beta}$
(it is also a factor algebra of the algebra $T(V)/\{I_-^q\}$).
Let us also set
$\oI^q_+=V^q_{2\alpha}$ and fix a
decomposition of the space $\vv=\oI^q_+\oplus \oI^q_-$ which
differs from
$\vv=I^q_+\oplus I^q_-$ but is a decomposition in the category $\uq$-Mod as
well (i.e., $\oI_{\pm}^q$ are objects of this category).

Note that due to the Kostant`s theorem (cf. \cite{LT}) the algebra
$T(V)/\{\oI_-\}$ where $\oI_{\pm}=\oI^1_{\pm}$ is the algebra of regular
 functions $Fun(\co)$
on the orbit $\co$ of a
 highest weight element $x$ in $V^*$.

\begin{proposition} \cite{DS1}
The deformation from $T(V)/\{\oI_-\}$ to $T(V)/\{\oI_-^q\}$ is flat.
\end{proposition}

Note that this statement was actually
proved in \cite{DGM} where
the algebra $T(V)/\{\oI_-\}$ was represented as a result of the deformational
 quantization of the bracket $\{\,\,,\,\,\}_R$.
In \cite{DG1} the R-matrix
bracket was introduced in the space of holomorphic sections of the linear
 bundles over a flag manifold
and it was quantized as well. More precisely, in the papers
 \cite{DGM}, \cite{DG1} it
was shown that the element $F\in \uq$, introduced by Drinfeld
(cf. \cite{D}) can be used as an intertwining operator between the
initial object and the quantized object.

Note that in \cite{S} the so-called quantum flag manifold was
treated in terms of projectivization of the h.w. orbits. In fact the
author  dealt with the algebras of the type $T(V)/\{\oI_-^q\}$ but did not
consider the quasiclassical limit.

Unfortunately, the element $F$ mentioned above cannot be used to quantize
 the Poisson pencil $\{\,\,,\,\,\}_{a,b}$ on any orbits of
R-matrix type (on symmetric orbits this series does not suit well
to quantize R-matrix brackets).
In \cite{DS2} another series $F$ quantizing the
Poisson pencil (2) on a Hermitian symmetric space in the sense of
deformation quantization was constructed.

Here we propose another way to define the q-deformation of some orbits in
$sl(n)^*$, and more generaly to quantize the Poisson pencil (2) on these
orbits (including the h.w. orbits).
We describe the orbit of an element $x\in \gggg^*$ by a system of equations
 and try to find its analog in the category $\uq$-Mod. The quotient algebra
by the ideal generated by the latter system is regarded as a
{\em quantum orbit}. Considering the corresponding quotient algebra of
algebra (3) we can quantize the whole Poisson pencil (2).

In the present paper we will restrict ourselves to a particular type of
orbits in $sl(n)^*$.

Let $C$ denotes the split Casimir, i.e., a generator of the trivial module
$V_0\subset \vv$ in the category $\gggg$-Mod.

We leave the proof of the following to the reader.
\begin{proposition}
Let $V=\gggg=sl(n)$. Consider the orbit $\co\in \gggg^*$ of the element
 $$x=M_a=diag(a,a,...,a, -(n-1)a), \,a\in k,\, a\not=0$$
 (we identify $\gggg$ with $\gggg^*$
 by means of the Killing form $<.,.>$).
Then the space $Fun(\co)$ is naturally identified with the quotient space
$$A_{0,1}^c=T(\gggg)/\{J_c\}\,\,{\rm where}\,\,
 J_c=\oplus_{\beta\not\in\{0,\, 2\alpha\}}\vb\oplus k(C-c),$$
for a constant $c=c(a)=<C,M_a\ot M_a>$ depending on the scalar $a$.
\end{proposition}

It is evident that the stabilizer of the point $x$ is $S(L(n-1)\times L(1))$
and therefore $\co=\sss$.

Note that the h.w. orbit in $\gggg^*$ is also defined by the same system of
 equations but with $c=0$ (actually the variety defined by the ideal
$\{J_{0}\}$ coincides with closure of the h.w. orbit over $k=\bC$).
Thus we have a family of orbits depending on the parameter $c\geq 0$.

Changing $\{J_c\}$ in this proposition to the ideal
$$\{\oplus_{\beta\not\in\{0,\, 2\alpha\}}
 (Im(id-h[\,\, ,\,\,])\vb)\oplus k(C-c)\}$$
we obtain a quotient space of the enveloping algebra $U(sl(n))$
(with the parameter $h$
introduced in the Lie bracket). We consider this algebra
 as quantization of the KKS bracket on the h.w. orbit $(c=0)$ or on a
 symmetric orbit $(c>0)$.

We will introduce a q-analog of this algebra in a natural way by setting
$$A_{h,q}^c=T(V)/\{\oplus_{\beta\not\in\{0,\, 2\alpha\}}
 k(Im(id-h[\,\, ,\,\,]_q)\vb^q)\oplus k(C_q-c)\}$$
where $C_q$ is ``the braided split Casimir'', i.e., a generator of the trivial
module $V_0^q$ in the category $U_q(sl(n))$-Mod
(we assume that $C_q\to C$ when $q\to 1$ or in other words
$C_q=C\,\,mod\,(q-1)$)\footnote{Note that the braided Casimir
differs from the so-called quantum Casimir which is an element of
the quantum group $\uq$.}.
This algebra depends on three parameters. The parameter $h$ arises from the
 quantization of the KKS bracket, $q$ is the braiding parameter
and $c$ labels the orbits.

The following proposition follows from the construction.

\begin{proposition}
Multiplication in the algebras $A_{h,q}^c$ is a $U_q(sl(n))$-morphism.
\end{proposition}

Note that in \cite{DG3} it was shown that the three-parameter deformation
from $A_{0,1}^0$ to
$A_{h,q}^c$ is flat in the case  $\gggg=sl(2)$.
It is very plausible that this deformation is flat in general. Our
optimism is based on a recent paper \cite{B} where the ``Koszulity'' of
the algebra $A^{0}_{0,1}$ is proved. We hope to discuss the problem of
the flatness elsewhere.

\section{Braided vector fields}

Thus we have constructed the
 three-parameter family of algebras $A_{h,q}^c$.
We consider the algebra $A_{0,q}^c$ as a q-analog of a function
algebra on the orbit $\co$. Discussing the problem of flatness, we
consider $h,\, q$ and $c$ as formal parameters. In this
Section we assume all parameters to be fixed, while the parameter $q$
is assumed to be generic.

Let us emphasize that all orbits of R-matrix type
are multiplicity-free, i.e., in the
decomposition of $Fun(\co)$ as a direct sum of
irreducible $\gggg$-modules, any two modules are pairwise non-isomorphic
(cf. \cite{GP}). For
the orbits under consideration, the space $Fun(\co)$
has the following decomposition
$$Fun(\co)=\oplus V_{n\alpha}\,\,{\rm where}\,\,
V_{n\alpha} \subset V^{\ot n},\,\,V=sl(n).$$

The space $A_{0,q}^c$ can be represented in a similar way as a direct sum
$$A_{0,q}^c=\oplus V_{n\alpha}^q\,\,{\rm with}\,\,
V_{n\alpha}^q \subset V^{\ot n}.$$

Let us consider the projectors
$$P^q_n : V^{\ot n} \to V^q_{n\alpha}$$
in the category $\uq$-Mod and introduce a ``braided'' analog of the coadjoint
vector fields in the following way.

For any element $u\in V$ introduce the operator
$$U^q:V\to V,\,\, U^qv=[u,v]_q,\,\, v\in V$$
and extend it to the spaces  $V_{n\alpha}^q$, setting
$$U^q_{(n)}v=n P^q_n(U^{q}\ot id_{n-1})v, \,\,v\in V_{n\alpha}^q\subset V^{\ot
n}$$
where $id_{n-1}$ is the identity operator in $V^{\ot (n-1)}$.

We also set $U^{q}_{(0)}1=0$.
 Thus any element $u\in V$ defines a family of operators
$$U^q_{(n)}: V_{n\alpha}^q\to V_{n\alpha}^q,\,\, n=0,1,2,...$$
which can be regaded as an operator $U^q:A_{0,q}^c \to A_{0,q}^c$.

It is easy to see that for $q=1$, the operator
$U=U^1$ coincides with the coadjoint
vector field generated by $u\in V$.
\begin{definition}
We call the operator
$U^q:A_{0,q}^c \to A_{0,q}^c$ a braided coadjoint vector field.
\end{definition}

Let $\{u_i\}$ be a base in $\gggg$ and
 $\{f_{\beta,i}(q)=\sum f_{\beta,i}^{k,l}(q)u_ku_l,
\,\,1\leq i\leq dim\,V_{\beta}^q\}$
 a base in the space $V_{\beta}^q \subset I_+^q$ (hereafter we omit the
$\ot$ sign).

Let us first consider the operators $U_{i}=U_{i}^{1}$ corresponding to the
elements $u_{i}$ in the non-deformed $(q=1)$ case.

The following proposition is obvious
\begin{proposition} On any orbit $\co$ where $x=M_a$ and on the h.w. orbit,
the operators $U_i$ satisfy the following equations
\begin{equation}
\sum f_{\beta,i}^{k,l}u_kU_l=0,\,\, \beta\not=2\alpha,\,\,
f_{\beta,i}^{k,l}=f_{\beta,i}^{k,l}(1).
\end{equation}
\end{proposition}

Note that this system of equations does not depend on the parameter $c=c(a)$.

Consider a left $A_{0,1}^c$-module
$$M=\{\sum
a^iU_i \,\,{\em mod}\,\, \sum
b^{\beta,i}f_{\beta,i}^{k,l}u_kU_l,\,\,a^i,b^{\beta,i}
\in A_{0,1}^c\}.$$
{}From the geometrical point of view this is the module of global sections
of the
 tangent bundle on the orbit $\co$.

\begin{definition}
The elements of the $A^c_{0,q}$-module
$$M_q=\{\sum
a^iU_i^q \,\, mod\, \sum b^{\beta,i}f_{\beta,i}^{k,l}(q)u_kU_l^q,\,
\,a^i,b^{\beta,i}
\in A_{0,q}^c\}$$
are called braided (tangent) vectors fields.
\end{definition}

To justify this definition we have to prove the following
\begin{conjecture}
Equations (4) hold if we
replace $U_i$ by $U^q_i$ and $f_{\beta,i}$ by $f_{\beta,i}(q)$.
\end{conjecture}
The case $\gggg=sl(2)$ was investigated in \cite{DG3} (cf. also
section 5).

\begin{remark}
Note that braided vector fields differ from those arising from
the quantum groups $\uq$. They look rather like super-vector fields.
If $A$ is an associative algebra equipped with
an involutive operator $S:A^{\ot 2}\to A^{\ot 2}$ satisfying the QYBE,
we say that it is $S$-commutative if $\mu S=\mu$ where $\mu$ is
the multiplication in $A$. For an S-commutative algebra it is natural to
introduce the notion of S-vector field using an ``S-analog'' of
the Leibniz rule $X\mu=\mu (X\ot id)(id+ S)$.

With regard to the algebras $A_{h,q}^{c}$ under consideration,
one can easily see (at least for the case when $c=0$) that they are
${\tilde S}$-commutative where
${\tilde S}$ is the involutive operator defined above.
It seems natural to introduce the q- (or braided) analog of vector
fields in such an algebra using the above  Leibniz rule but with
$\tilde S$ instead of $S$.
However this definition of the
q-deformed vector fields is not consistent.
In our aproach we do not use any form of the Leibniz rule.

Note that our definition of q-deformed vector fields is close
ideologically to  the ``braided partial
derivatives'' introduced by Sh. Majid \cite{M1}. However since the deformation
from $T(V)/\{I_-\}$ to $T(V)/\{I_-^q\}$ is not flat it is hopeless to
get a flat deformation of the module of vector fields, while
 it seems very plausible that
the module $M_q$ of braided vector fields introduced above
represents a flat deformation of
the initial module.
\end{remark}

\section{Poisson brackets connected to quantum orbits}

In this section we discuss the quasiclassical counterpart of the quantum orbits
considered above and we investigate the relations between
the brackets $\{\,\,,\,\,\}_{KKS}$ and $\{\,\,,\,\,\}_r^{red}$ on the symmetric
 orbits in $\gggg^*$.

First we introduce the R-matrix bracket on any homogeneous
$G$-space $G/H$ (or more generally on any manifold $M$ equipped with a
 representation $\rho:\gggg\to Vect(M)$) as follows
$$\{f,g\}_R=\mu<\rho^{\ot 2}(R), df\ot dg>.$$
Here $\mu:Fun(M)^{\ot 2}\to Fun(M)$ is the ordinary commutative
multiplication in the space $Fun(M)$ and $R$ is the R-matrix (1).

\begin{definition}
We say that the space $M$ is of R-matrix type if this bracket is Poisson.
\end{definition}

Consider the element $\varphi\in \wedge^3(\gggg)$ defined as follows
\begin{equation}
\varphi=[R^{12}, R^{13}]+[R^{12}, R^{23}]+[R^{13}, R^{23}].
\end{equation}
Note that $\varphi$ is $G$-invariant and that it is
a generator of the space $H_3(\gggg)$.
It is evident that a homogeneous $G$-space $M$ is of R-matrix type iff
$\varphi$
 restricted to $M$ is equal to 0 in the following sense:
$$\mu<\rho^{\ot 3}(\varphi),\,\,df\ot dg \ot dh>=0$$
 for any $f,\,g,\, h\in Fun(M)$.

Hence, the property of a variety
to be of R-matrix type does not depend on a choice
of a modified R-matrix (we say that an element
$R\in\wedge^2(\gggg)$ is a modified R-matrix if the r.h.s. of
(5) is $G$-invariant).

In  \cite{DGM} it was proved that a h.w. orbit in any $G$-module $V$ is of
R-matrix type. In \cite{DG2} the same property was proved for
 symmetric spaces.
And in \cite{GP} all orbits in $\gggg^*$ (over the field
$k=\bC$)  were classified.
\begin{proposition}
(cf. \cite{GP})
1. If $\co\in\gggg^*$ is of R-matrix type then $x\in \gggg^*$ is either
semisimple or nilpotent.\\
2. If $x$ is semisimple then $\co$ is of R-matrix type iff $\co$ is symmetric.
\end{proposition}

Now consider the brackets $\{\,\,,\,\,\}_{r,l}$ defined in Section 1.
Using the fact that the element $\varphi$ is $G$-invariant it is easy to see
 that the brackets $\{\,\,,\,\,\}_R=\{\,\,,\,\,\}_l$ and
$\{\,\,,\,\,\}_r^{red}$ become Poisson simultaneously.

 Thus on any symmetric
space the bracket $\{\,\,,\,\,\}_r^{red}$
(if it is well-defined) is Poisson and $G$-invariant. If
$M=\co$ is an orbit in $\gggg^*$ there exists another $G$-invariant bracket,
namely the KKS bracket. So it is natural to pose the
question of the relation between brackets $\{\,\,,\,\,\}_{KKS}$ and
$\{\,\,,\,\,\}_r^{red}$.

Although we deal mainly  with the non-compact real
form of the group $G$ we will consider also the compact form
of the group $G$ and the algebra $\gggg$, which we denote
by $G_c$ and $\gggg_c$ respectively. Note that $R_c=\sqrt{-1}R$ where
 $R$ is the R-matrix (1) is an element of
$\wedge^2(\gggg_c)$. Thus we can introduce all the above brackets on the
homogeneous $G_c$-spaces and introduce the notion of orbit of R-matrix type.

Consider an element $x\in \gggg$ belonging to the Cartan subalgebra of the
algebra $\gggg$. Then $\sqrt{-1}x\in\gggg_c$. Note that for such an
element $x$ the reduced bracket $\{\,\,,\,\,\}_r^{red}$ is well-defined.

It is clear that the orbits
$\co\in\gggg^*$ and ${\cal O}_{\sqrt{-1}x}\in\gggg_c^*$ are
symmetric simultaneously. Assume that they are symmetric.

Then the
 orbit ${\cal O}_{\sqrt{-1}x}$ is a symmetric Hermitian space and therefore
$dim\,H^2({\cal O}_{\sqrt{-1}x},\, {\bf C})=1$. Using this fact one can
show that the brackets KKS and $\{\,\,,\,\,\}_r^{red}$ are proportional to each
other both for orbits ${\cal O}_{\sqrt{-1}x}\in \gggg_c^*$
and $\co\in\gggg^*$
(this proof was first given in \cite{DG2}). Here we will give another proof
using an explicit construction from \cite{KRR}.

If $\co\in \gggg^{*}$ is a symmetric orbit of an element $x$ belonging to the
Cartan subalgebra, the isotropy subalgebra ${\frak h}$ of the point $x$
can be described in a way similar to \cite{KRR}
(although in \cite{KRR} the compact form
is considered but actually we are  dealing only with
local properties which do not depend on the chosen real form of $\gggg$).

Let $\alpha_1,...\alpha_{n}$ be the set of all
simple roots of the Lie algebra $\gggg$ and $\alpha_k$ be one of them.
Consider the algebra ${\frak h}$ generated by the Cartan subalgebra and
elements
$\{X_{\alpha},\,X_{-\alpha}\}$
such that in the decomposition $\alpha=\sum a_i\alpha_i,\,\,\,a_k=0$.
Then according to the Cartan classification of
the Hermitian symmetric
spaces we can assume that the isotropy subalgebra has this form with
 arbitrary simple root $\alpha_k$ for the series $A_{n}$ or with  some
 special root for the simple algebras of other series
(cf. \cite{KRR} where all possible $\alpha_k$ are indicated).

Set $\overline\Delta=\{\alpha\in\Delta_+,\,
 X_{\alpha}\not \in {\frak h}\}$ (recall that $\Delta_+$ denotes the set of all
 positive roots). Then the KKS bracket being at point
$x$ has the form
$$\{f,g\}_{KKS}=\mu<\rho^{\ot 2}_r(\frac{1}{2}\sum_{\alpha\in \overline\Delta}
\frac{1}{\alpha (x)}X_{\alpha}\wedge X_{-\alpha}),\,df\ot dg>.$$

Let us show that $\alpha(x)$ does not depend on
$\alpha \in \overline\Delta$.
Indeed if $\alpha \in \overline\Delta$ then
$\alpha=\alpha_k+\sum_{i\not= k}a_i\alpha_i$ where $\alpha_{i}$
is a root such that $X_{\alpha_i}\in{\frak h}$.
Since $[x,\,X_{\alpha_i}]=\alpha_i(x)X_{\alpha_i}=0$ we have
$\alpha_{i}(x)=0$.
Therefore $\alpha(x)=\alpha_{k}(x)+\sum_{i\not=k}a_{i}\alpha_{i}(x)=
\alpha_{k}(x)=const$.

Thus at point $x$ the brackets $\{\,\,,\,\,\}_{KKS}$ and
 $\{\,\,,\,\,\}_r^{red}$
are proportional to each other. Now using the fact that
 both brackets are $G$-invariant we can state the following

\begin{proposition}
Let us consider a symmetric
 orbit $\co=G/H\in\gggg^*$ of an element $x$ belonging to the Cartan
 subalgebra of $\gggg$.
Then the brackets $\{\,,\,\,\}^{red}_r$ and $\{\,,\,\,\}_{KKS}$
 coincide on $M=\co$ up to a factor.
\end{proposition}

On the h.w. orbit, the bracket $\{\,,\,\,\}^{red}_r$ is also
well-defined but it is identically equal to zero. So the brackets
 $\{\,,\,\,\}_{R}$ and $\{\,,\,\,\}_{SD}^{red}$ coincide on such an
orbit.
The ''quantum pairs'' method is no longer valid to quantize this type of
orbits, while our approach applies in principle to any orbit of R-matrix type.

We now want to discuss the relations between the above algebras
$A_{h,q}^{c}$ and
the Poisson pencil (2)
(all parameters here are assumed to be formal).

\begin{proposition} (\cite{D}) There exists a series
 $F\in \ug^{\ot 2}[[\nu]]$ such that
the following conditions are satisfied:\\
1. $F=1 \,\, mod\,\nu$;\\
2. there exists a Hopf algebra isomorphism
$$i: \ug[[\nu]]\to\uq,\,\,i=id+\nu i_1+\nu^2 i_2+...$$
between the algebra $\ug[[\nu]]$
equipped with the coproduct $\de_F(a)=F^{-1}\de(a)F$ and an
appropriate antipod and  the quantum group $\uq$ (here $q=e^{\nu}$);\\
3. the universal quantum R-matrix ${\cal R}$ corresponding to
the quantum group $\uq$ can be written in the form
${\cal R}=(F^{-1})^{21} e^{\nu \frac{C}{2}}F$, where $C$ is the split Casimir
in $\gggg^{\ot 2}$.
\end{proposition}

Using the last property one can deduce that
\begin{equation}
F-F^{21}=\nu R \,\,mod\,\nu^2
\end{equation}
 where $R$ is the R-matrix (1).

It is easy to see that the operator
${\tilde S}:\vv\to\vv$ introduced above can be represented in the
following form
$$
{\tilde S}=\tilde{\rho}_q^{\ot 2}(F^{-1})\sigma\tilde{\rho}_q^{\ot 2}(F)
$$
where $\tilde{\rho}_q=\rho_q\,i$,
$i: \ug[[\nu]]\to\uq$ is the above isomorphism and
$\rho_q :\uq\to End(\vv)$ is the representation of
the quantum group $\uq$ defined in Section 2.

Assume now that the deformation from $A^{0}_{0,1}$ to $A^{c}_{h,q}$ is flat.
Then, using relation (6) and the fact that $\tilde{\rho}_q$ tends to
$\rho$ when
$q\to 1$ we can state that the Poisson bracket corresponding to the family of
 algebras $A^c_{0,q}$ is $\{\,\,,\,\,\}_R$ (up to a factor) on the orbits
under consideration.
This implies that the quasiclassical limit of the family $A^c _{h,q}$
coincides with the Poisson pencil (2).

\section{Example: the case of $sl(2)$}

 In this section we
set  $V=\gggg=sl(2)$.

It is well-known that the quantum group $\uq$ is generated by elements
 $\{H,X,Y\}$  satisfying the relations
$$[H,X]=2X,\; [H,Y]=-2Y,\; [X,Y]=\frac{q^H-q^{-H}}{q-q^{-1}}.$$
A coproduct in this algebra can be defined in different ways, for example
 as follows
$$\de(X)=X\ot 1+q^{-H}\ot X,\; \de(Y)=1\ot Y +Y\ot q^H,\;
\de(H)=H\ot 1+ 1\ot H$$
but various definitions give rise to isomorphic Hopf algebras (equipped
with the appropriate antipodes).

Let us now decompose the space $V^{\ot 2}$ as a direct sum of
irreducible $\uq$-modules:
$\vv=V_0^q\oplus V_1^q \oplus V_2^q$ ($i$ is the spin of the space $V_i^q$). To
 describe this decomposition  more explicitly, we fix a base
 $\{u,v,w\}$ in $V$ and define the action of the quantum qroup $\uq$ on $V$ as
 follows
$$Hu=2u,\ Hv=0,\ Hw=-2w,\ Xu=0,\ Xv=-\qq u,\ Xw=v,$$
$$Yu=-v,\ Yv=\qq w,\ Yw=0.$$
Then the spaces $V_i^q$ are
$V_0^q=kC_{q}$,
$$V_1^q={\rm Span}(q^2u v - v u,\ (q^3+q)( u w-w u) +(1-q^2)v v,\
 -q^2v w + w v),$$
$$V_2^q={\rm Span}(uu,\ uv+q^2vu,\  uw-q vv+q^4wu,\
vw+q^2 wv,\ ww),$$
where $C_q=(q^3+q)uw+vv+(q+q^{-1})wu$
is the q-deformed split Casimir.

Let us introduce  now the q-Lie bracket in $\gggg$ arising in
the above construction.
\begin{proposition} \cite{DG3} The multiplication table for the bracket
$[\,\,,\,\,]_q$ in the case of $\gggg=sl(2)$ is
$$[u,u]_q=0,\ [u,v]_q=-q^2Mu,\ [u,w]_q=\qq^{-1}Mv,$$
$$[v,u]_q=Mu,\ [v,v]_q=(1-q^2)Mv,\ [v,w]_q=-q^2Mw,$$
$$[w,u]_q=-\qq^{-1}Mv,\ [w,v]_q=Mw,\ [w,w]_q=0,$$
where $M\in k$ is an arbitrary factor.
\end{proposition}

Consider the algebra $A_{h,q}^c$ defined above.
It has the following form
$$A_{h,q}^c=T(V)/\{J_c^q\},\, J_c^q=Span(C_q-c,\,\,
q^2u v - v u + 2hu,$$
$$ (q^3+q)( u w-w u) +(1-q^2)v v- 2hv,\,\,
-q^2v w + w v- 2hw).$$

Let us define the operators $U^{q},\,V^{q},\,W^{q}$ as above, setting
$$U^{q}u=0,\, U^{q}v=-q^2 Mu,\, U^{q}w=(q+q^{-1})^{-1}Mv, ...,
W^{q}v=Mw,\,W^{q}w=0.$$

According to the above scheme these operators extended to the higher
components are regarded as the
 coadjoint vector fields. It is easy to demonstrate by a direct computation
that these fields satisfy the
 following relation
$$f=(q^3+q)uW^{q}+vV^{q}+(q+q^{-1})wU^{q}=0.$$

Thus the $A_{0,q}^c$-module of the braided tangent vector fields can be
 introduced in this particular case as follows
$$M_q=\{aU+bV+cW,\, \,{\rm mod}\, df,\,\,a,b,c,d \in \ac\}.$$

{\bf Acknowledgements.} This paper was mainly prepared in Centre de
 Math\'ematiques de l'Ecole
Polytechnique where one of the authors (D.G.) enjoyed a one-year position
and the other (J.D.) was a short-time visitor. We are grateful
to the Centre de Math\'ematiques for hospitality.
We also wish to thank
 B.Enriquez, A.Guichardet, S.Khoroshkin and Y.Kosmann--Schwarzbach
 for stimulating discussions.

\end{document}